\documentclass[conference]{IEEEtran}
\IEEEoverridecommandlockouts
\usepackage{cite}
\usepackage{amsmath,amssymb,amsfonts}
\usepackage{algorithmic}
\usepackage{graphicx}
\usepackage{textcomp}
\usepackage{xcolor}
\usepackage{float}
\usepackage{url}
\usepackage{authblk}
\usepackage[T1]{fontenc}
\usepackage[polish]{babel}
\usepackage[utf8]{inputenc}
\usepackage[caption=false,font=footnotesize]{subfig}

\def\BibTeX{{\rm B\kern-.05em{\sc i\kern-.025em b}\kern-.08em
    T\kern-.1667em\lower.7ex\hbox{E}\kern-.125emX}}

\begin{document}

\title{Image recognition based on optical spike processing with exciton-polaritons}

\author[1]{Olgierd Jeziorski}
\author[2]{Jakub Rogala}
\author[1]{Krzysztof Tyszka}

\affil[1]{Faculty of Physics, University of Warsaw}
\affil[2]{Interdisciplinary Centre for Mathematical and Computational Modelling, University of Warsaw}

\maketitle

\section{Introduction}
Neuromorphic computing systems aim to mimic the way a biological brain processes information \cite{Tyszka2023, Opala2024}, offering advantages over classical von Neumann architectures such as parallel, event-driven computation and greater energy efficiency for certain tasks\cite{Opala2024}.
One of the major neuromorphic paradigms is Spiking Neural Networks (SNNs), which process information through discrete spikes, with information encoded in spike timing and frequency \cite{Matuszewski21, Maass1999}. By processing information in parallel and asynchronously, these systems offer energy efficiency and speed advantages over classical neural networks\cite{Opala2024, Matuszewski21}.
Despite this, electronic implementations of SNNs face limitations in spike transmission speeds, which are typically restricted to the microsecond range \cite{Young2019}. Realizing SNNs in fully optical systems offers a potential solution, promising sub-ns processing speeds, even lower energy consumption, and parallelism \cite{Tyszka25}. While optical systems lack the versatility and fine-tuning capabilities of digital platforms, they remain beneficial for applications requiring rapid neural processing.
Several attempts have been made to create fully optical neural processing units on platforms such as VCSELs \cite{Hurtado2012}, phase-change-material (PCM)  cells \cite{Feldmann19}, DFB lasers \cite{Zhang2024}, and nanoscale resonant-tunneling diodes (nanoRTDs) \cite{Romeira2023}.
A fundamental requirement for any artificial neural network is the presence of nonlinear activation functions, which are essential for learning complex, non-separable data patterns. Without nonlinearity, multi-layered networks mathematically collapse into a single linear transformation, severely limiting their computational capability. However, realizing strong, energy-efficient optical nonlinearity remains a major bottleneck in the development of fully optical SNNs, as standard photonic platforms are inherently linear.
One of the new candidates for fully optical neuromorphic platforms, which is highly promising to bridge this gap is exciton-polaritons in semiconductor microcavities \cite{Sanvitto2016, Fraser2016}. These bosonic quasiparticles, which emerge due to strong coupling between photons confined in a microcavity and excitons confined within a semiconductor quantum well, can form a non-equilibrium Bose-Einstein condensate at room temperature. Polaritons inherit their very low effective mass from their photonic component, enabling propagation speeds up to a few percent of the speed of light and correspondingly fast switching times, while their excitonic component provides strong optical nonlinearities at power thresholds orders of magnitude lower than those of conventional optical media \cite{Deng2010, Carusotto2013, Opala2018}. These properties have already enabled the realization of basic logic elements such as switches \cite{Genco2025, Fraser_2017}, transistors \cite{Ballarini2013, Zasedatelev2019}, and gates \cite{Li2024}, as well as proof-of-concept machine learning implementations, including a polariton-based classifier shown to outperform linear classification algorithms \cite{Ballarini2020, Opala2019}.

It has been demonstrated that exciton-polaritons can replicate a biological neuron's information processing in the optical domain, following the Leaky Integrate-and-Fire (LIF) model, with ultrafast pulsing and high energy efficiency \cite{Tyszka2023}. Specifically, sub-ns reaction times and energy efficiency below 1 pJ per pulse have been shown \cite{Tyszka2023, Tyszka25}. This provides additional motivation to further study the use of exciton-polaritons for processing information encoded in pulses, especially given the recent demonstration of room-temperature exciton-polariton neural networks in perovskite crystals \cite{Opala2024}. Several networks exploiting the nonlinearity of the condensate thresholding mechanism have already been developed\cite{Opala2024, Matuszewski21, Opala22}. However, these implementations largely exploit spatial or steady-state nonlinearities of the condensate, leaving the computational potential of its purely temporal, dynamics unexplored.

Here, we take advantage of the pulse-processing capabilities of polariton condensates \cite{Tyszka2023} to construct a fully temporal spiking neural network: information is encoded exclusively in the timing and amplitude of optical pulses driving a single-mode condensate, with no spatial degrees of freedom involved. The condensate's nonlinear thresholding mechanism, which reproduces the dynamics of a Leaky Integrate-and-Fire neuron, is exploited directly through its time evolution, transforming temporally encoded input into a higher-dimensional feature space, which is then sampled to obtain feature vectors on which linear classification can be performed. To benchmark this mechanism against standard approaches, we evaluate it on a temporally encoded, downscaled MNIST dataset, comparing against both a linear (logistic regression) and a nonlinear (feed-forward network) baseline. The use of MNIST dataset provides a widely recognized reference point. This work can be viewed as a proof of concept, that purely temporal polariton dynamics can perform beneficial nonlinear computations for classification tasks.

\section{Model and Dynamics}
\subsection{The Single-Mode Approximation}
The spatial and temporal dynamics of an exciton-polariton condensate are typically described by the open-dissipative Gross-Pitaevskii equation coupled with a rate equation for the reservoir. However, under conditions of strong spatial confinement or spatially uniform pumping, spatial variations can be neglected. We utilize a single-mode approximation that treats the condensate as a zero-dimensional system, focusing purely on its temporal evolution.

\subsection{Coupled Rate Equations}
The model describes the temporal evolution of three interdependent state variables:
\begin{itemize}
    \item \textbf{Condensate Population ($n_C$):} The population density of condensing polaritons.
    \item \textbf{Active Reservoir ($n_R$):} Particles available to scatter directly into the condensate (e.g., excitons in close vicinity).
    \item \textbf{Inactive Reservoir ($n_I$):} High-energy particles created directly by the non-resonant pump.
\end{itemize}

Interactions are governed by scattering rates and decay lifetimes. The inactive reservoir feeds the active reservoir at a scattering rate $\kappa$. The active reservoir feeds the condensate at a rate $R$. Scattering into the condensate is stimulated by the existing condensate population ($R n_R n_C$), providing the nonlinear thresholding behavior essential for signal generation. The system of coupled ordinary differential equations is defined as:
\begin{align}
\label{eq:ode1} \frac{dn_I}{dt} &= P(t) - \kappa n_I^2 - \gamma_I n_I, \\
\label{eq:ode2} \frac{dn_R}{dt} &= \kappa n_I^2 - R n_R n_C - \gamma_R n_R, \\
\label{eq:ode3} \frac{dn_C}{dt} &= R n_R n_C - \gamma_C n_C.
\end{align}
Here, $P(t)$ represents the external optical pump. 
% The constant $\epsilon \approx 10^{-6}$ accounts for intrinsic quantum or thermal fluctuations required to seed the condensation process.

\section{Data Encoding and Methodology}

\subsection{Temporal Pulse Encoding}
The MNIST dataset consists of 60000 greyscale, 28x28 pixel images of hand-drawn digits from 0 to 9. To speed up the process of model training and optimisation as well as making the dataset less linearly separable we used a modified MNIST dataset downscaled to $10 \times 10$ pixels ($M = 100$), containing $1000$ elements per class. The models were evaluated on two distinct variations of this dataset: the standard 10-class digit dataset and a challenging 2-class subset containing only the digits ``7'' and ``9''. This binary subset was specifically selected because these digits exhibit highly correlated structural shapes, making them the most difficult for standard linear regression models to reliably separate and learn without nonlinear transformations.

To interface with the time-domain condensate model, we transform each image into a sequence of Gaussian pulses constructing the pump function $P(t)$:
\begin{equation}
P(t) = \sum_{j=1}^{M} A_j \exp\left(- \frac{(t - t_j)^2}{2\sigma^2}\right)
\end{equation}
The temporal position of each pulse is $t_j = j \cdot \Delta t$. The variance parameter $\sigma$ relates to the pulse temporal duration via:
\begin{equation}
    \sigma = \frac{t_{\text{FWHM}}}{2\sqrt{2\ln2}}
\end{equation}
where $t_{\text{FWHM}}$ represents the full width at half maximum of the optical pump pulse. The amplitude $A_j$ is linearly mapped from the corresponding pixel intensity $I_j \in [0, 1]$:
\begin{equation}
A_j = A_{\text{min}} + I_j (A_{\text{max}} - A_{\text{min}})
\end{equation}

\subsection{Sampling and Feature Extraction}The total simulation interval $T_{\text{total}}$ is divided into $N$ equal integration windows (bins). The discrete feature value $v_k$ for the $k$-th bin is obtained by integrating the condensate density over that window: $v_k = \int_{t_{k-1}}^{t_k} n_C(t) dt$ for $k \in \{1, 2, \dots, N\}$, where $t_k = \frac{k T_{\text{total}}}{N}$. These extracted values are scaled to achieve a zero mean and unit variance before being passed to a multinomial logistic regression model.

\subsection{Bayesian Optimization of the Parameter Space}
To establish an optimal baseline, Bayesian optimization was performed on the pulse encoding parameters (separation $\Delta t$, pulse width $t_{\text{FWHM}}$, minimum amplitude $A_{\text{min}}$, and maximum amplitude $A_{\text{max}}$). This optimization was conducted under fixed, physically realistic values for the condensate transition parameters outlined in Table~\ref{tab:parameters}.

Through Bayesian optimization, we identified an optimal set of pulse parameters: separation $\Delta t \approx 30\ [\mathrm{ps}]$, width $\approx 5\ [\mathrm{ps}]$, $A_{\text{min}} \approx 10\ [\mathrm{a.u.}]$, and $A_{\text{max}} \approx 30\ [\mathrm{a.u.}]$. Using these encodings and $N=100$ integration windows, we evaluated the system against standard benchmarks.

\begin{table}[H]
\caption{Physical Condensate (ODE) Parameters}
\begin{center}
\begin{tabular}{|c|c|c|}
\hline
\textbf{Parameter} & \textbf{Description} & \textbf{Value} \\
\hline
$\gamma_C$ & Condensate decay rate & $1 /12 \ \mathrm{ps}^{-1}$ \\
\hline
$\gamma_R$ & Active reservoir decay rate & $5 \cdot 10^{-3} \ \mathrm{ps}^{-1}$ \\
\hline
$\gamma_I$ & Inactive reservoir decay rate & $10^{-3} \ \mathrm{ps}^{-1}$ \\
\hline
$\kappa$ & Scattering rate (Inactive to Active) & $5\cdot10^{-2}$ \\
\hline
$R$ & Stimulated scattering rate & $6\cdot10^{-4}$ \\
\hline
\end{tabular}
\label{tab:parameters}
\end{center}
\end{table}

\subsection{Operational Heatmaps}
To understand the physical limits of the system's nonlinear processing ability, we mapped performance across the broader parameter space. By fixing two parameters at their optimized values and varying the remaining two, we generated heatmaps to locate boundaries of effective processing.

\begin{figure}[H]
\centerline{\includegraphics[width=0.45\textwidth]{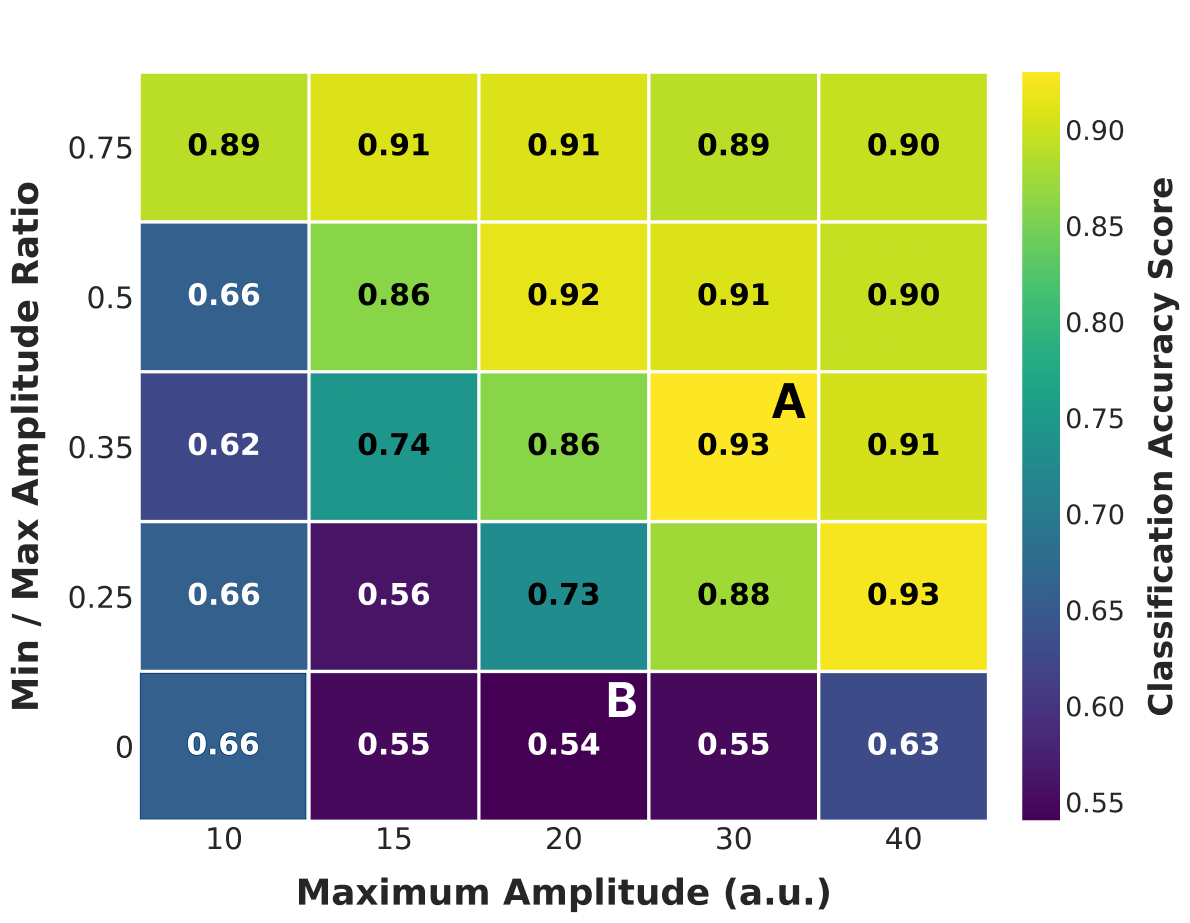}}
\caption{Classification accuracy as a function of maximum pulse amplitude and minimum amplitude. A distinct diagonal band of high accuracy ($\sim$92--93\%) dictates the optimal operational regime.}
\label{fig:heatmap_amp}
\end{figure}

Fig.~\ref{fig:heatmap_amp} shows a diagonal region of optimal performance. Performance degrades outside this regime due to two failure modes:
\begin{itemize}
    \item \textbf{Under-stimulation:} When amplitudes are too low, the condensate is not stimulated frequently enough for macroscopic condensation, preventing nonlinear integration.
    \item \textbf{Over-stimulation (Saturation):} Excessively high amplitudes prevent the condensate from relaxing to its ground state. The system processes information linearly, bypassing the computational advantage of the node.
\end{itemize}

\begin{figure}[H]
\centerline{\includegraphics[width=0.45\textwidth]{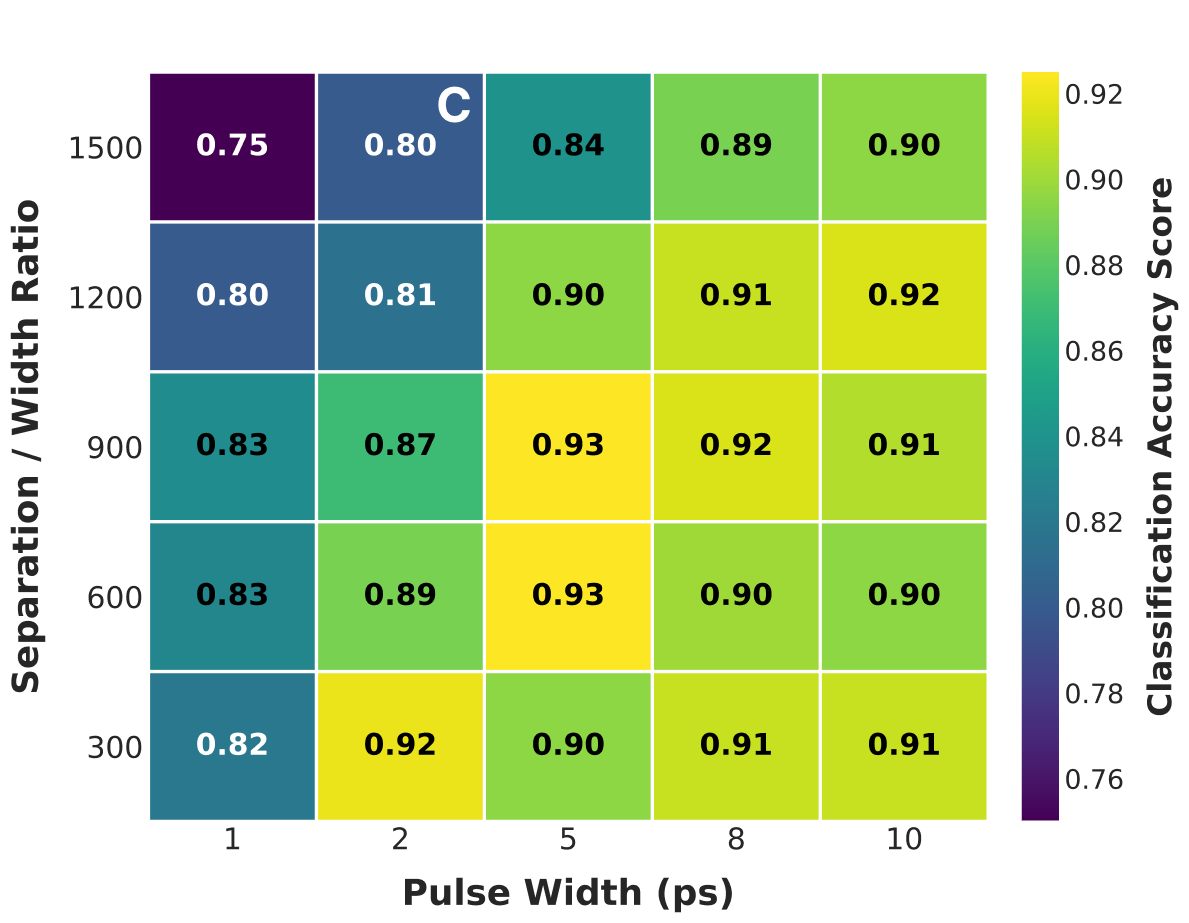}}
\caption{Classification accuracy as a function of pulse width and separation. The optimal regime balances stimulation with relaxation time.}
\label{fig:heatmap_time}
\end{figure}

A similar boundary dictates temporal encoding (Fig.~\ref{fig:heatmap_time}). Narrow pulses with large separation cause under-stimulation, while overly broad pulses with short separation suppress nonlinear transient dynamics through constant stimulation.

\subsection{Operational Regimes and Temporal Dynamics}
To physically illustrate the performance boundaries identified in the parameter heatmaps, we analyzed the temporal dynamics and classification confusion matrices across three distinct operational regimes: optimal, under-stimulated, and saturated, labeled A, B, and C respectively.

% Fig.~\ref{fig:combined_regimes} consolidates these observations. For each corresponding parameter combination marked on the earlier heatmaps, the figure displays the classification confusion matrix alongside the pump sequence $P(t)$ and the condensate response $n_C(t)$ for pairs of sample digits. 

In regime A (optimal), the condensate exhibits strong, nonlinear spiking that accurately captures the temporal features of the input while allowing for sufficient relaxation. The corresponding confusion matrix confirms that these optimal parameters yield balanced, high-accuracy predictions across all classes. 

Conversely, in regime B (under-stimulated), combining a low maximum amplitude with a minimum amplitude of zero results in a weakly driven system. The resulting condensate response is several orders of magnitude lower than in the optimal regime, yielding erratic and noisy signals that fail to produce useful computational features.

Similarly, regime C (isolated pulses) fails to leverage the system's nonlinear interaction capabilities. Although individual pulse amplitudes are high enough to trigger condensation, the narrow pulse widths and wide temporal separations prevent consecutive pulses from interacting. Because each pulse independently triggers condensation without cumulative temporal integration, the system defaults to extracting purely linear features. Consequently, long-range pixel interactions cannot be captured, and performance defaults to standard linear limits.

% \begin{figure}[H]
% \centerline{\includegraphics[width=0.55  \textwidth]{macierze_pomylek.png}}
% \label{fig:combined_regimes}
% \end{figure}% Figure A: Optimal regime
\begin{figure}[H]
    \centering
    \includegraphics[width=0.35\textwidth]{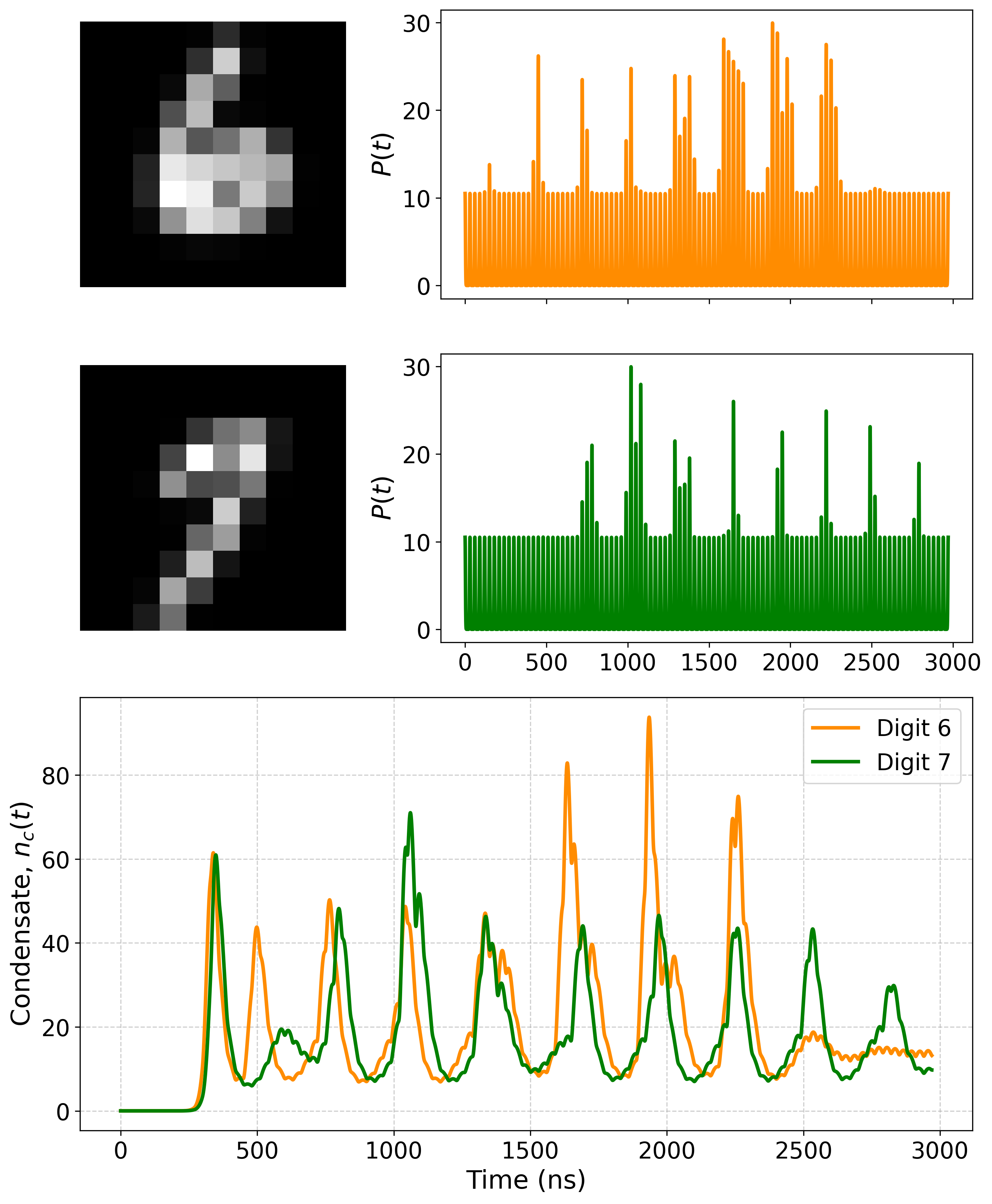}
    \caption{%
        \textbf{Optimal regime.}
        Example pump pulses $P(t)$ and corresponding condensate
        responses $n_C(t)$ in the optimal operating regime.
    }
    \label{fig:regime_A}
\end{figure}

% Figure B: Under-stimulated regime
\begin{figure}[H]
    \centering
    \includegraphics[width=0.35\textwidth]{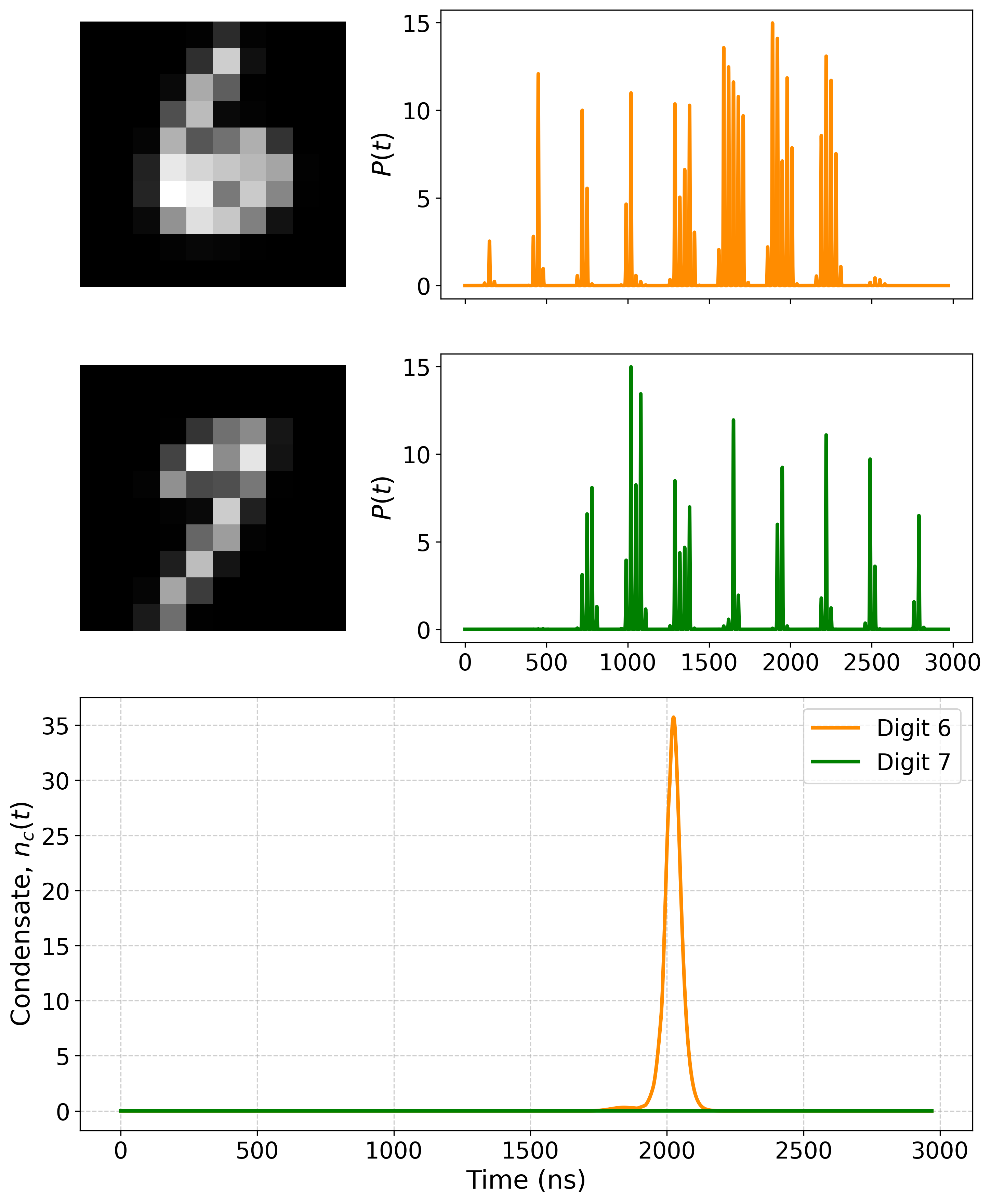}
    \caption{%
        \textbf{Under-stimulated regime.}
        Example pump pulses $P(t)$ and corresponding condensate
        responses $n_C(t)$ when the system is insufficiently stimulated.
    }
    \label{fig:regime_B}
\end{figure}

% Figure C: Pulses too far apart
\begin{figure}[H]
    \centering
    \includegraphics[width=0.35\textwidth]{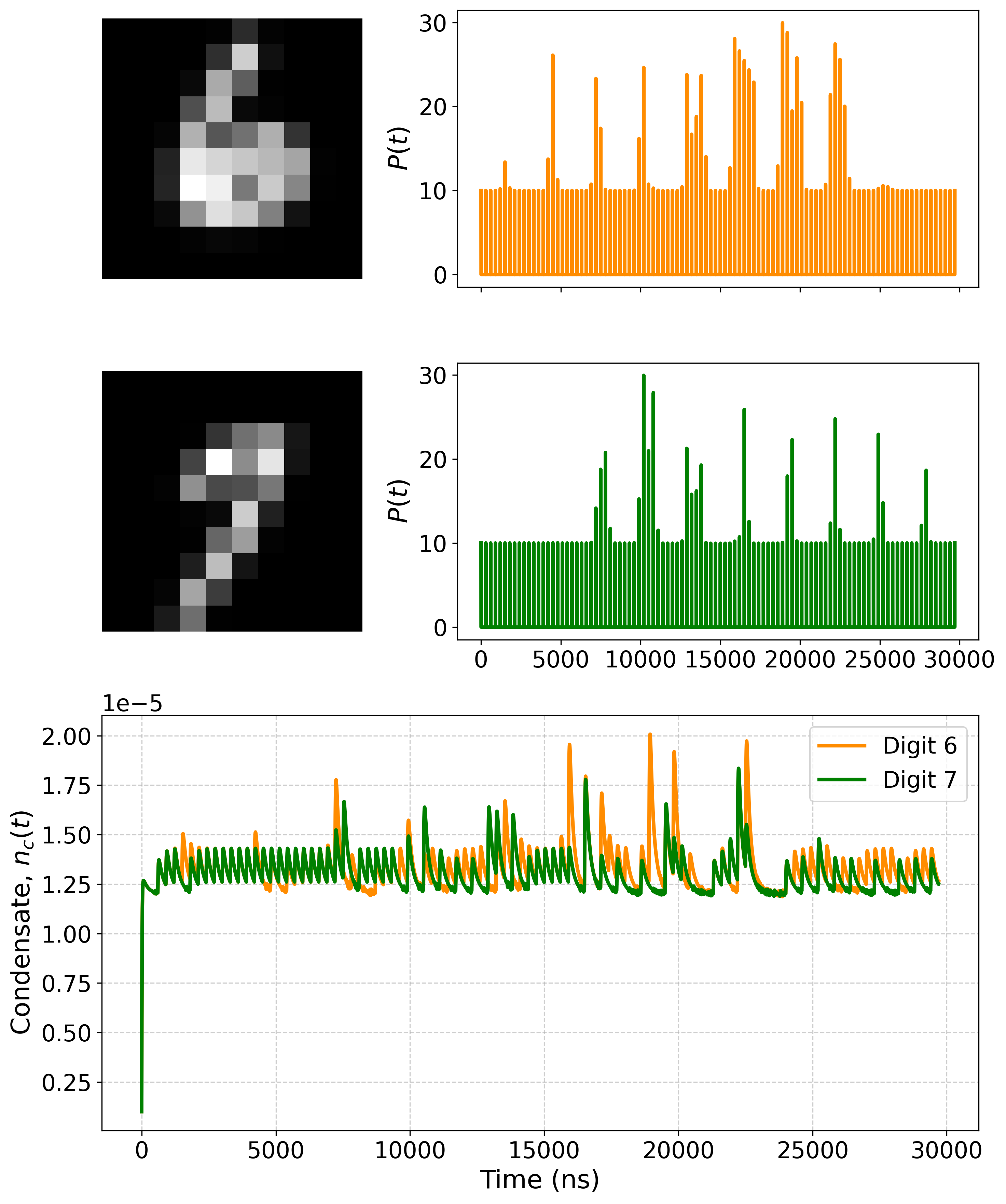}
    \caption{%
        \textbf{Pulses too far apart.}
        Example pump pulses $P(t)$ and corresponding condensate
        responses $n_C(t)$ when the temporal separation between pulses
        is too large.
    }
    \label{fig:regime_C}
\end{figure}
\section{Results and Discussion}
\subsection{Classification Performance}
\begin{figure}[t]
\centerline{\includegraphics[width=0.45\textwidth]{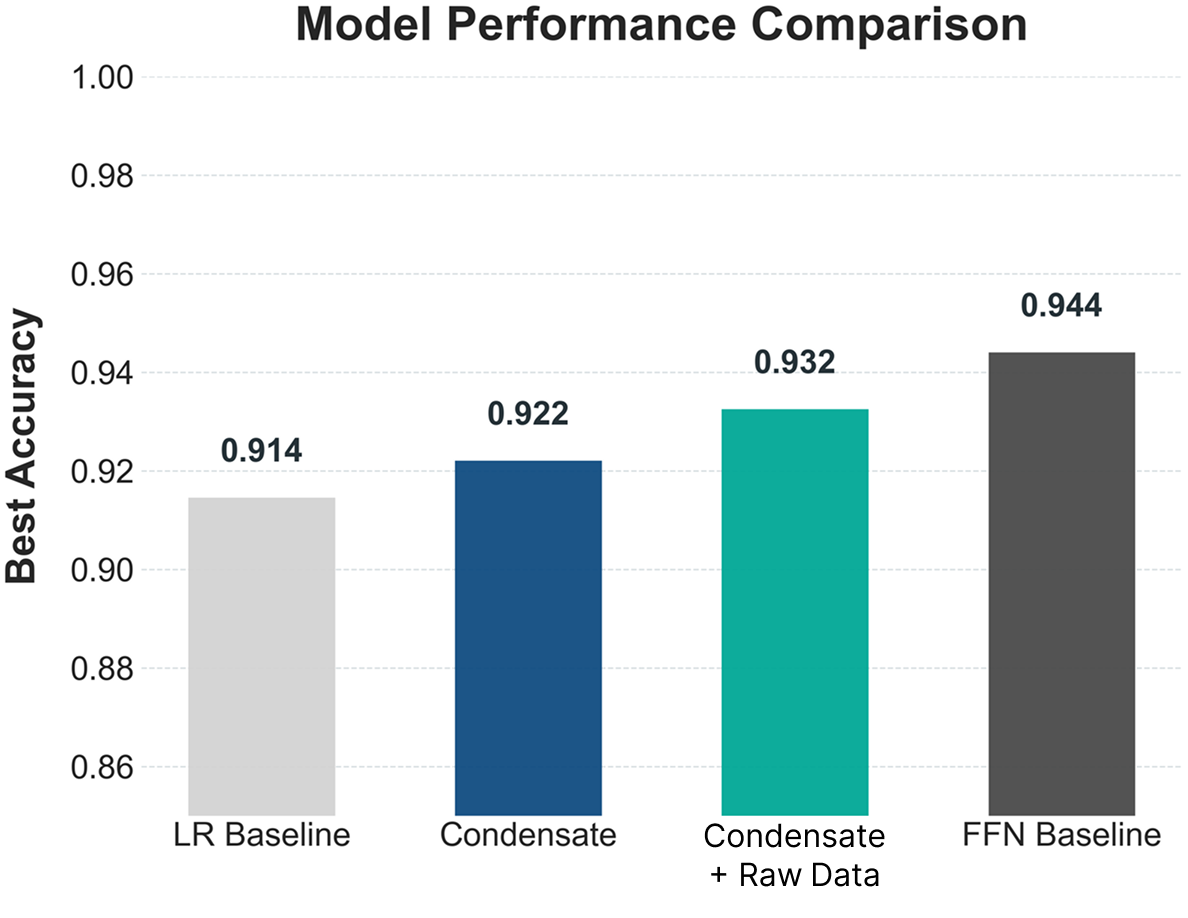}}
\caption{Classification accuracy comparison between the base linear model, pure condensate model, condensate + raw pixel model, and FFN.}
\label{fig:barchart}
\end{figure}

Fig.~\ref{fig:barchart} compares the classification accuracy across four distinct models: a base linear model trained on raw pixels, a model using pure condensate features, a combined model using condensate features concatenated with raw pixels, and a Feed-Forward Network (FFN).

In the combined model, the $N$-dimensional temporal feature vector extracted from the condensate is concatenated directly with the original $M$-dimensional raw pixel array, extending the input vector passed to the multinomial logistic regression classifier. This combination is designed to explicitly test whether the polariton node introduces higher-order relational metrics that are missing from raw linear observations.

The standalone condensate features achieve an accuracy of approximately 92.2\%, matching and slightly surpassing the $\sim$91\% baseline of the pure linear model. When the $N$-dimensional nonlinear temporal features are concatenated with the raw pixel data, the combined system achieves an accuracy of 93.2\%. 

This performance enhancement can be framed as a reservoir computing projection or a physical manifestation of the classic kernel trick. The exciton-polariton condensate maps the low-dimensional, linearly inseparable raw pixel data into a higher-dimensional temporal feature space governed by \eqref{eq:ode1}--\eqref{eq:ode3}. In this expanded state space, the boundaries between highly correlated and structurally similar classes become more distinct and linearly separable, allowing the subsequent linear regression layer to draw superior classification boundaries.

\subsection{Temporal Dynamics and Sampling Resolution}
\begin{figure}[H]
\centerline{\includegraphics[width=0.45\textwidth]{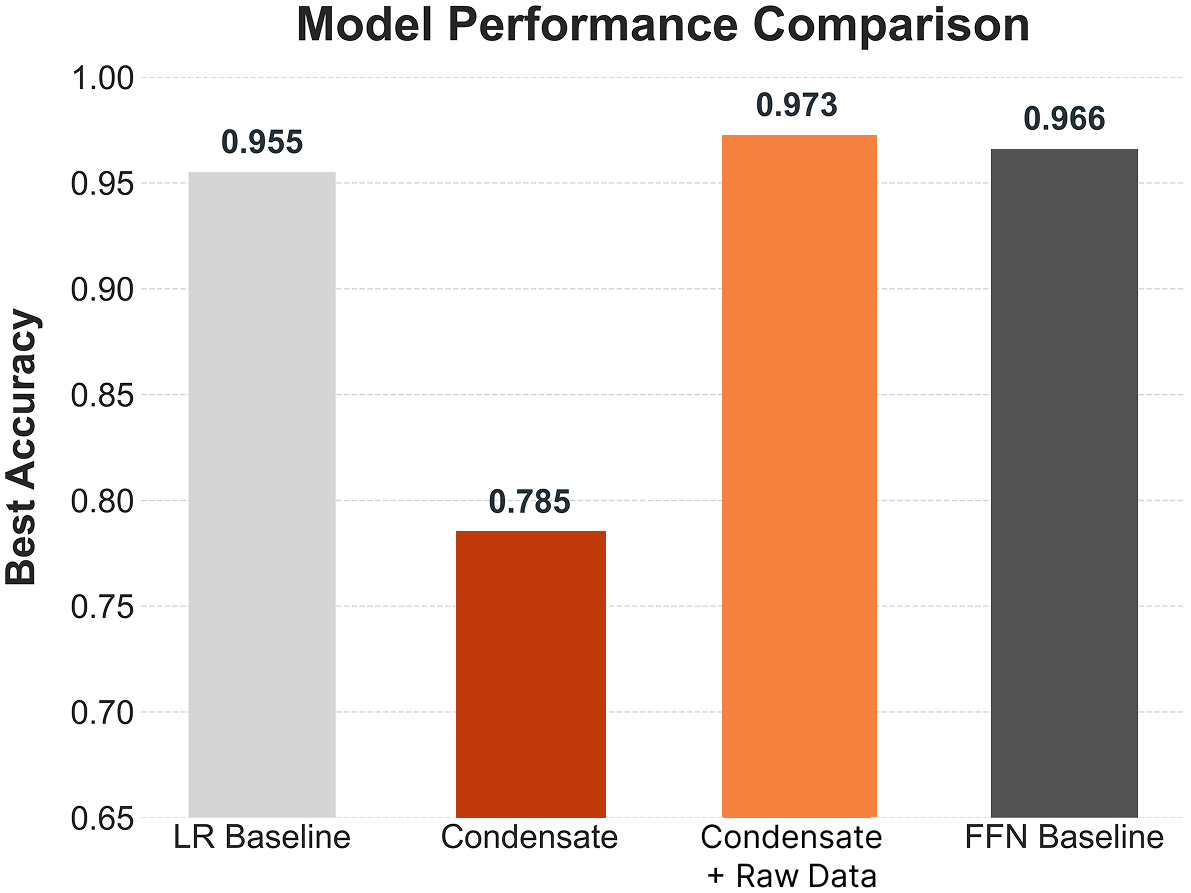}}
\caption{Classification accuracy comparison between the base linear model, pure condensate model, condensate + raw pixel model, and FFN specifically for the challenging 2-digit binary task.}
\label{fig:barchart_2digit}
\end{figure}

We evaluated the models strictly on the challenging 2-digit binary subset. Fig.~\ref{fig:barchart_2digit} compares the accuracies of the base linear model, the pure condensate model, the combined (condensate + raw pixels) model, and an FFN for this specific task. The combined architecture shows a dramatic improvement in accuracy over the baseline, reinforcing the conclusion that the condensate effectively extracts critical higher-dimensional nonlinear features that linear models cannot naturally extract.

\begin{figure}[H]
\centerline{\includegraphics[width=0.45\textwidth]{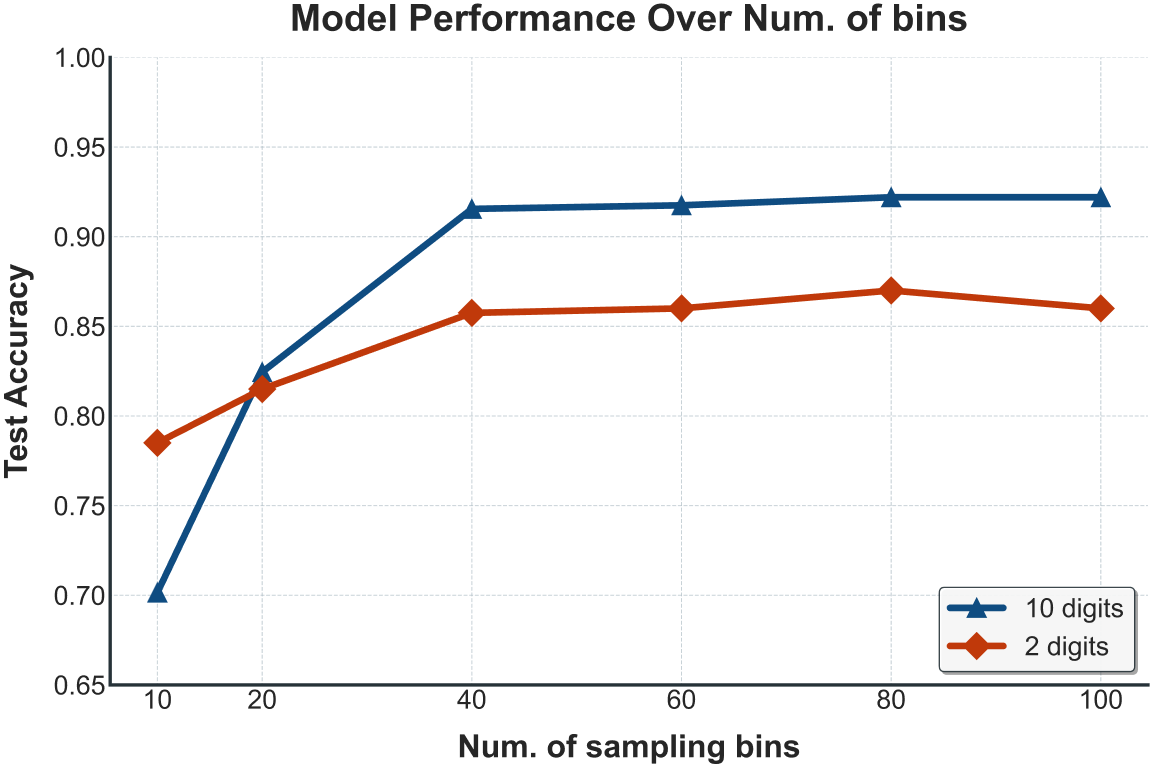}}
\caption{Model accuracy as a function of the number of integration bins ($N$) utilizing pure condensate features for both the 10-digit and 2-digit binary tasks.}
\label{fig:n_plot_pure}
\end{figure}

With this nonlinear advantage established, we analyzed the impact of sampling resolution on system performance for both the standard 10-digit task and the 2-digit subset. Fig.~\ref{fig:n_plot_pure} displays the performance trajectories for models trained exclusively on pure condensate features. In both tasks, accuracy scales with an increasing number of integration bins and plateaus near $N=40$. Experimentally, this indicates that extracting standalone information from the condensate requires a higher temporal resolution (e.g., a fast streak camera) to capture the fine details of the high-speed optical dynamics.

\begin{figure}[H]
\centerline{\includegraphics[width=0.45\textwidth]{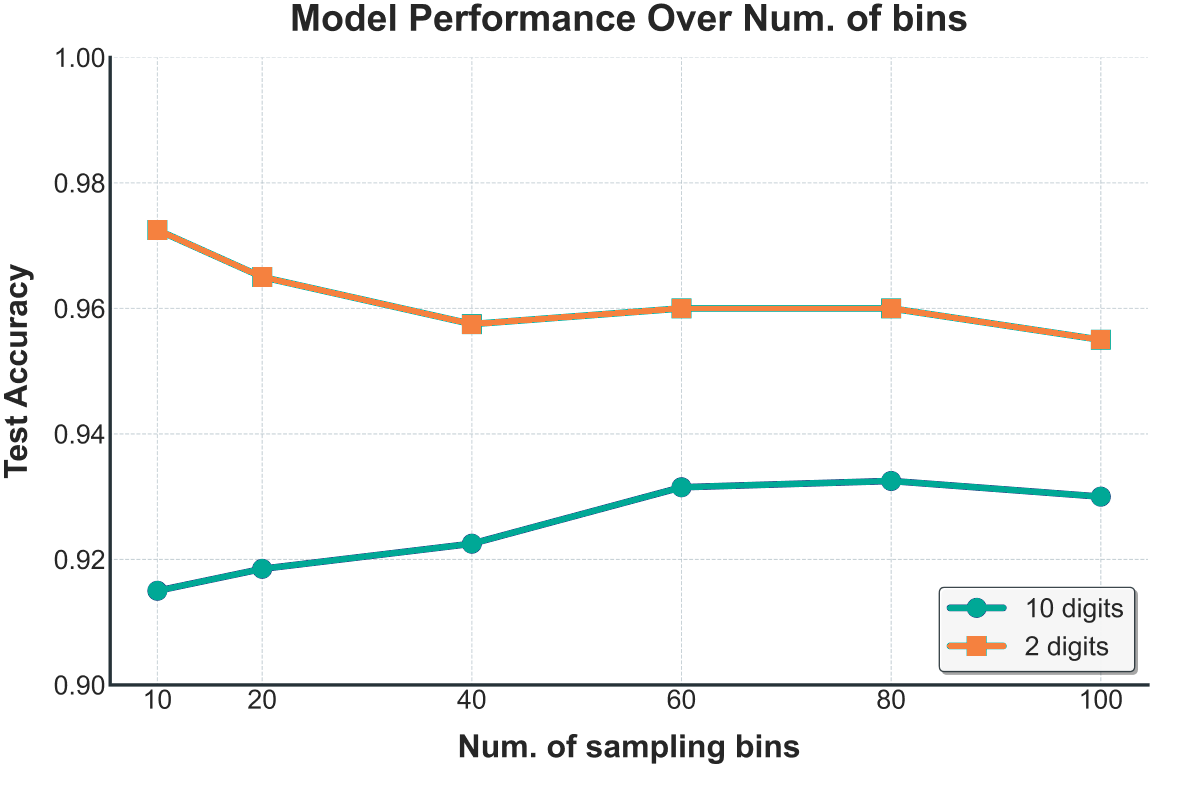}}
\caption{Model accuracy as a function of the number of integration bins ($N$) utilizing combined (condensate + raw pixel) features for both the 10-digit and 2-digit binary tasks.}
\label{fig:n_plot_combined}
\end{figure}

Fig.~\ref{fig:n_plot_combined} illustrates the classification performance when raw pixel data is combined with the condensate features. Remarkably, the accuracy remains stable and high across all tested values of $N$ for both the 10-digit and 2-digit tasks. This highlights a critical practical advantage: while capturing the fine details of standalone linear processing requires high temporal resolution, the most beneficial nonlinear processing occurs over longer, macroscopic timescales. As a result, the system can utilize the condensate's powerful nonlinear properties without strictly mandating ultra-fast sampling rates or expensive high-speed readout hardware.

\section{Conclusion}
The simulation of exciton-polariton condensate driven by temporally encoded image sequences demonstrates its viability as a practical optical neuromorphic platform. The simulated condensate successfully performs beneficial nonlinear computations, transforming spatial data into higher-dimensional temporal features.

When applied to a temporally encoded MNIST classification task, the system achieved a baseline accuracy of 92.2\% using only the transient dynamics of the optical condensate. Combining these polaritonic temporal features with raw pixel data increased the classification accuracy to 93.2\%. While the MNIST dataset is largely linearly separable by itself and therefore not ideal for fully showcasing the advantages of a nonlinear framework, this improvement confirms that the polaritonic system introduces useful, higher-order information processing that complements standard linear processing. Because the current benchmark did not heavily demand complex data separation, evaluating this architecture on datasets better suited for nonlinear tasks will likely yield even better results.

Furthermore, analysis of the integration windows shows that this nonlinear advantage remains consistent across different sampling resolutions. The system extracts key macroscopic features without requiring ultra-fast sampling rates or specialized readout hardware.

Ultimately, these findings demonstrate that exciton-polariton condensates function as effective platforms for neuromorphic computing. By combining stron nonlinear effects with fast physical dynamics, polaritonic systems offer a clear pathway toward developing high-speed, stable, and fully optical information processing devices.

\bibliographystyle{unsrt}
\bibliography{bibliography}

@article{Tyszka2023,
author = {Tyszka, Krzysztof and Furman, Magdalena and Mirek, Rafał and Król, Mateusz and Opala, Andrzej and Seredyński, Bartłomiej and Suffczyński, Jan and Pacuski, Wojciech and Matuszewski, Michał and Szczytko, Jacek and Piętka, Barbara},
title = {Leaky Integrate-and-Fire Mechanism in Exciton–Polariton Condensates for Photonic Spiking Neurons},
journal = {Laser \& Photonics Reviews},
volume = {17},
number = {1},
pages = {2100660},
doi = {https://doi.org/10.1002/lpor.202100660},
url = {https://onlinelibrary.wiley.com/doi/abs/10.1002/lpor.202100660},
eprint = {https://onlinelibrary.wiley.com/doi/pdf/10.1002/lpor.202100660},
year = {2023}
}

@misc{Opala2024,
      title={Room temperature exciton-polariton neural network with perovskite crystal}, 
      author={Andrzej Opala and Krzysztof Tyszka and Mateusz Kędziora and Magdalena Furman and Amir Rahmani and Stanisław Świerczewski and Marek Ekielski and Anna Szerling and Michał Matuszewski and Barbara Piętka},
      year={2024},
      eprint={2412.10865},
      archivePrefix={arXiv},
      primaryClass={physics.optics},
      url={https://arxiv.org/abs/2412.10865}, 
}

@article{Matuszewski21,
  title = {Energy-Efficient Neural Network Inference with Microcavity Exciton Polaritons},
  author = {Matuszewski, M. and Opala, A. and Mirek, R. and Furman, M. and Kr\'ol, M. and Tyszka, K. and Liew, T.C.H. and Ballarini, D. and Sanvitto, D. and Szczytko, J. and Pi\ifmmode \mbox{\k{e}}\else \k{e}\fi{}tka, B.},
  journal = {Phys. Rev. Appl.},
  volume = {16},
  issue = {2},
  pages = {024045},
  numpages = {16},
  year = {2021},
  month = {Aug},
  publisher = {American Physical Society},
  doi = {10.1103/PhysRevApplied.16.024045},
  url = {https://link.aps.org/doi/10.1103/PhysRevApplied.16.024045}
}

@article{Opala22,
  title = {Training a Neural Network with Exciton-Polariton Optical Nonlinearity},
  author = {Opala, A. and Panico, R. and Ardizzone, V. and Pi\ifmmode \mbox{\k{e}}\else \k{e}\fi{}tka, B. and Szczytko, J. and Sanvitto, D. and Matuszewski, M. and Ballarini, D.},
  journal = {Phys. Rev. Appl.},
  volume = {18},
  issue = {2},
  pages = {024028},
  numpages = {8},
  year = {2022},
  month = {Aug},
  publisher = {American Physical Society},
  doi = {10.1103/PhysRevApplied.18.024028},
  url = {https://link.aps.org/doi/10.1103/PhysRevApplied.18.024028}
}

@article{Feldmann19,
    title = {All-optical spiking neurosynaptic networks with self-learning capabilities},
	volume = {569},
	journal = {Nature},
	author = {Feldmann, J. and Youngblood, N. and Wright, C. D. and Bhaskaran, H. and Pernice, W. H. P.},
	month = may,
	year = {2019},
	pages = {208--214},
}

@article{Tyszka25,

  title = {Advancing Optical Spiking Neural Networks with Exciton-Polaritons},

  author = {Tyszka, K. and Opala, A. and Pi\ifmmode \k{e}\else \k{e}\fi{}tka, B.},

  journal = {J. Phys. Photonics},

  volume = {7},

  issue = {4},

  pages = {041002},

  year = {2025},

  month = {Oct},

  publisher = {IOP Publishing},

  doi = {10.1088/2515-7647/ae0aa0},

  url = {https://doi.org/10.1088/2515-7647/ae0aa0}

}

@article{Young2019,
author = {Young, Aaron and Dean, Mark and Plank, James and Rose, Garrett},
year = {2019},
month = {09},
pages = {135606-135620},
title = {A Review of Spiking Neuromorphic Hardware Communication Systems},
volume = {7},
journal = {IEEE Access},
doi = {10.1109/ACCESS.2019.2941772}
}

@article{Hurtado2012,
author = {Hurtado, A. and Schires, K. and Henning, I. D. and Adams, M. J.},
year = {2012},
month = {03},
pages = {103703},
title = {Investigation of vertical cavity surface emitting laser dynamics for neuromorphic photonic systems},
volume = {100},
journal = {Applied Physics Letters},
doi = {10.1063/1.3692726}
}

@book{Maass1999,
editor = {Maass, Wolfgang and Bishop, Christopher M.},
year = {1999},
title = {Pulsed Neural Networks},
publisher = {MIT Press}
}

@article{Ballarini2013,
	title = {All-optical polariton transistor},
	volume = {4},
	doi = {10.1038/ncomms2734},
	journal = {Nature Communications},
	author = {Ballarini, D. and De Giorgi, M. and Cancellieri, E. and Houdré, R. and Giacobino, E. and Cingolani, R. and Bramati, A. and Gigli, G. and Sanvitto, D.},
	year = {2013},
	pages = {1778}
}

@article{Zasedatelev2019,
author = {Zasedatelev, Anton V. and Baranikov, Anton V. and Urbonas, Darius and Scafirimuto, Fabio and Scherf, Ullrich and Stöferle, Thilo and Mahrt, Rainer F. and Lagoudakis, Pavlos G.},
year = {2019},
month = {06},
pages = {378-383},
title = {A room-temperature organic polariton transistor},
volume = {13},
number = {6},
journal = {Nature Photonics},
doi = {10.1038/s41566-019-0392-8}
}

@article{Li2024,
author = {Li, Hui and Chen, Fei and Jia, Haoyuan and Ye, Ziyu and Zhou, Hang and Luo, Song and Shi, Junheng and Sun, Zhenrong and Xu, Huailiang and Xu, Hongxing and Byrnes, Tim and Chen, Zhanghai and Wu, Jian},
year = {2024},
pages = {864-869},
title = {All-optical temporal logic gates in localized exciton polaritons},
volume = {18},
journal = {Nature Photonics},
doi = {10.1038/s41566-024-01483-2}
}

@article{Genco2025,
author = {Genco, Armando and others},
year = {2025},
month = {07},
pages = {6490},
title = {Femtosecond switching of strong light-matter interactions in microcavities with two-dimensional semiconductors},
volume = {16},
journal = {Nature Communications},
doi = {10.1038/s41467-025-61607-2}
}

@article{Fraser_2017,
doi = {10.1088/1361-6641/aa730c},
url = {https://doi.org/10.1088/1361-6641/aa730c},
year = {2017},
month = {aug},
publisher = {IOP Publishing},
volume = {32},
number = {9},
pages = {093003},
author = {Fraser, Michael D},
title = {Coherent exciton-polariton devices},
journal = {Semiconductor Science and Technology}
}

@article{Ballarini2020,
author = {Ballarini, D. and Gianfrate, A. and Panico, R. and Opala, A. and Ghosh, S. and Dominici, L. and Ardizzone, V. and De Giorgi, M. and Lerario, G. and Gigli, G. and Liew, T. C. H. and Matuszewski, M. and Sanvitto, D.},
year = {2020},
month = {05},
pages = {3506--3512},
title = {Polaritonic Neuromorphic Computing Outperforms Linear Classifiers},
volume = {20},
journal = {Nano Letters},
doi = {10.1021/acs.nanolett.0c00435}
}

@article{Opala2019,
  title = {Neuromorphic Computing in Ginzburg-Landau Polariton-Lattice Systems},
  author = {Opala, Andrzej and Ghosh, Sanjib and Liew, Timothy C.H. and Matuszewski, Micha\l{}},
  journal = {Phys. Rev. Appl.},
  volume = {11},
  issue = {6},
  pages = {064029},
  numpages = {10},
  year = {2019},
  month = {Jun},
  publisher = {American Physical Society},
  doi = {10.1103/PhysRevApplied.11.064029},
  url = {https://link.aps.org/doi/10.1103/PhysRevApplied.11.064029}
}

@article{Deng2010,
  title = {Exciton-polariton Bose-Einstein condensation},
  author = {Deng, Hui and Haug, Hartmut and Yamamoto, Yoshihisa},
  journal = {Rev. Mod. Phys.},
  volume = {82},
  issue = {2},
  pages = {1489--1537},
  numpages = {0},
  year = {2010},
  month = {May},
  publisher = {American Physical Society},
  doi = {10.1103/RevModPhys.82.1489},
  url = {https://link.aps.org/doi/10.1103/RevModPhys.82.1489}
}

@article{Carusotto2013,
  title = {Quantum fluids of light},
  author = {Carusotto, Iacopo and Ciuti, Cristiano},
  journal = {Rev. Mod. Phys.},
  volume = {85},
  issue = {1},
  pages = {299--366},
  numpages = {0},
  year = {2013},
  month = {Feb},
  publisher = {American Physical Society},
  doi = {10.1103/RevModPhys.85.299},
  url = {https://link.aps.org/doi/10.1103/RevModPhys.85.299}
}

@article{Opala2018,
  title = {Theory of relaxation oscillations in exciton-polariton condensates},
  author = {Opala, Andrzej and Pieczarka, Maciej and Matuszewski, Micha\l{}},
  journal = {Phys. Rev. B},
  volume = {98},
  issue = {19},
  pages = {195312},
  numpages = {9},
  year = {2018},
  month = {Nov},
  publisher = {American Physical Society},
  doi = {10.1103/PhysRevB.98.195312},
  url = {https://link.aps.org/doi/10.1103/PhysRevB.98.195312}
}

@article{Sanvitto2016,
  title={The road towards polaritonic devices.},
  author={Daniele Sanvitto and St{\'e}phane K{\'e}na‐Cohen},
  journal={Nature materials},
  year={2016},
  volume={15 10},
  pages={
          1061-73
        },
  url={https://api.semanticscholar.org/CorpusID:30199439}
}

@article{Fraser2016,
author = {Fraser, Matthew D. and Höfling, Sven and Yamamoto, Yoshihisa},
year = {2016},
month = {09},
pages = {1049--1052},
title = {Physics and Applications of Exciton-Polariton Lasers},
volume = {15},
journal = {Nature Materials},
doi = {10.1038/nmat4762}
}

@article{Romeira2023,
author = {Romeira, Bruno and Adão, Ricardo and Nieder, Jana B. and Al-Taai, Qusay and Zhang, Weikang and Hadfield, Robert H. and Wasige, Edward and Hejda, Matěj and Hurtado, Antonio and Malysheva, Ekaterina and others},
year = {2023},
month = {07},
pages = {033001},
title = {Brain-inspired nanophotonic spike computing: challenges and prospects},
volume = {3},
journal = {Neuromorphic Computing and Engineering},
doi = {10.1088/2634-4386/acdf17}
}

@article{Zhang2024,
author = {Zhang, Yuna and Xiang, Shuiying and Yu, Chengyang and Gao, Shuang and Han, Yanan and Guo, Xingxing and Zhang, Yahui and Shi, Yuechun and Hao, Yue},
title = {Photonic Neuromorphic Pattern Recognition with a Spiking DFB-SA Laser Subject to Incoherent Optical Injection},
journal = {Laser \& Photonics Reviews},
volume = {19},
number = {1},
pages = {2400482},
doi = {https://doi.org/10.1002/lpor.202400482},
url = {https://onlinelibrary.wiley.com/doi/abs/10.1002/lpor.202400482},
eprint = {https://onlinelibrary.wiley.com/doi/pdf/10.1002/lpor.202400482},
year = {2025}
}

\end{document}